\documentclass[11pt]{article}

\pdfoutput=1

\usepackage{euscript}
\usepackage{amssymb}
\usepackage{amsfonts}
\usepackage{amsbsy}
\usepackage{epsfig}
\usepackage{amsthm}
\usepackage{amscd}
\usepackage{amstext}
\usepackage{verbatim}
\usepackage{amsmath}
\usepackage{cancel}
\usepackage{capt-of}
\usepackage{empheq}
\usepackage{subfigure}
\usepackage{color}

\usepackage[pdftex]{hyperref}

\def\ben{\begin{equation}}
\def\een{\end{equation}}

\def\be{\begin{equation}}
\def\ee{\end{equation}}
\def\beq{\begin{equation}}
\def\eeq{\end{equation}}
\def\ba{\begin{array}}
\def\ea{\end{array}}

\def\dalemb#1#2{{\vbox{\hrule height .#2pt
       \hbox{\vrule width.#2pt height#1pt \kern#1pt
               \vrule width.#2pt}
       \hrule height.#2pt}}}

\newcommand{\bea}{\begin{eqnarray}}
\newcommand{\eea}{\end{eqnarray}}

\definecolor{dyellow}{rgb}{1.,0.8,.0}
\definecolor{myblue}{rgb}{.1,.1,.7}
\definecolor{dcyan}{rgb}{.0,.6,.6}
\definecolor{cyan}{rgb}{0.4,1.0,1.0}
\definecolor{dmagenta}{rgb}{0.6,0.0,0.6}
\definecolor{brown}{rgb}{0.6,0.2,0.}
\definecolor{darkblue}{rgb}{.0,.0,0.5}
\definecolor{darkred}{rgb}{0.75,0.0,0.0}
\definecolor{orange}{rgb}{1.,.6,.0}
\definecolor{dorange}{rgb}{0.8,.4,.0}
\definecolor{green}{rgb}{0.0,1.0,0.0}
\definecolor{darkgreen}{rgb}{0.0,0.6,0.0}
\definecolor{purple}{rgb}{.4,.0,.4}
\definecolor{lightgrey}{rgb}{0.7, 0.7, 0.7}
\definecolor{grey}{rgb}{0.4, 0.4, 0.4}

\def\blue{\color{blue}}

\begin{document}

\begin{center}

{ \Large {\bf
Zeroth Order Phase Transition in a Holographic Superconductor with Single Impurity
}}

\vspace{1cm}

Hua Bi Zeng$^{1}$ and Hai-Qing Zhang$^{2}$

\vspace{1cm}

{\small
$^{1}${ School of Mathematics and Physics, Bohai University
JinZhou 121000, China }}


{\small
$^{2}${Institute for Theoretical Physics, Utrecht University, \\Leuvenlaan 4, 3584 CE Utrecht, The Netherlands}}

\vspace{1.6cm}

\end{center}

\begin{abstract}

We investigate the single normal impurity effect in a superconductor by the holographic method. When the size of impurity is much smaller than the host superconductor, we can reproduce the Anderson theorem, which states that a conventional s-wave superconductor is robust to a normal (non-magnetic) impurity with small impurity strength. However, by increasing the size of the impurity in a fixed-size host superconductor, we find a decreasing critical temperature $T_c$ of the host superconductor,  which
agrees with the results in condensed matter literatures. More importantly, the phase transition at the critical impurity strength (or the
critical temperature) is of zeroth order.

\end{abstract}

\pagebreak

\section{Introduction}

 Duality between a large $N$ $d$-dimensional strongly coupled quantum field
theory and a $(d+1)$-dimensional  classical gravity theory (the AdS/CFT correspondence)\cite{adscft}
has become a very powerful tool to study the condensed
matter phenomena\cite{r1,r2,r3,r4}. In particular,  a black hole background coupled to a charged
 scalar theory was constructed in \cite{h1} to study the holographic superconductor. In that paper, the author found that in the probe limit, there is a
critical temperature $T_c$ below which the bosonic operator of the boundary field theory
has a finite expectation value, which corresponds to a homogeneous
s-wave superconductor.  Reviews of the holographic superconductor
can be found in \cite{r2,r5,r6}. In this paper we will extend this homogeneous
construction to a single normal impurity effect{\footnote{{The normal impurity is a substitution atom without magnetism, but
with different electron configuration from the host superconductor.}}} in a holographic superconductor,
in which the order parameter becomes spatially dependent due to the impurity.
Other studies of inhomogeneous holographic superconductors can be found in \cite{p0,p1,p2,p3,p4,p5,p6,p7,p8,p9,p10,p11,GarciaGarcia:2012zd}.
 {In this paper, the numerical technique is roughly following} \cite{p0,p1}.

To study a superconductor with an impurity substitution is important in order to understand superconductivity in condensed matter physics, for reviews see \cite{BVZ}.
Early important experimental results show that the conventional superconductivity is robust
to small concentrations of normal impurity, especially a single normal impurity.
These results can be understood by the Anderson's theorem \cite{anderson}.
In which, Anderson found that at the mean field level with a small impurity concentration, the gap equation
keeps the same if the gap is still uniform and the density of the states is unchanged compared to the
case without an impurity. Thus the critical temperature $T_c$ remains to $T_{c0}$, which is the critical temperature of the pure host superconductor.
Anderson's theorem is however an approximate statement, in fact even there is only a small impurity,
the local properties of the impurity will change a lot \cite{gapconfig1,gapconfig2}.
In these two papers, the order parameter of a superconductor in the presence of a single impurity
was obtained by solving the self-consistent
Bogoliubov-de Gennes (BdG) equations. Although the host condensate will not be affected by
the impurity, the condensate at the impurity is suppressed a lot.
Hence, one can naturally expect that if the size of the impurity is increased, the
host superconductor properties will change as well. This phenomenon requires us to
study the single impurity effect on different length scales,  from
lattice spacing to coherence length, even to the host superconductor size. Specifically, when the
impurity size is of lattice spacing, or in other words, in the limit of the localization size, the host superconductor will keep the same
as the pure case\cite{ma};
When the impurity size approaches to the coherence length, which is smaller than the host superconductor, properties of
the host sample will begin to change; However, if we keep increasing the impurity
size to the host superconductor size, superconductivity are expected to reduce substantially\cite{ghosal,xiang}.

The interesting question is how to understand the single impurity effect from AdS/CFT correspondence.
In this paper, we construct a gravity dual of a superconductor with a normal impurity
in the center of the superconducting host. We reproduce the Anderson theorem
that  $T_c$ of the host superconductor will not be affected by the impurity
when the size of the impurity  is smaller compared to the host; However, we find that it does reduce the gap at the impurity site as studied in \cite{gapconfig1,gapconfig2}.
This contradiction to Anderson theorem can be understood since Anderson theorem is an approximate
statement about the thermodynamic average of the system in the mean field theory level, while we are solving the whole spatially dependent
gap equations in the gravity  which corresponds to strongly coupled field theory.  For a larger size impurity , we find that $T_c$ decreases
dramatically and finally the impurity can destroy superconductivity as the temperature  $T$  of the
host superconductor is close to $T_{c0}$.

The paper is arranged as follows: In Section \ref{sect:reviewbcs} we review the known results for a
host superconductor with a normal impurity in the center;
In Section \ref{sect:anderson} we set up the model holographically; The numerical results of the suppression of the superconductivity can be found in Section \ref{sect:suppresion}; We draw our conclusions and discussions in Section \ref{sect:conclusion}.

\section{Brief review of  the normal impurity effect in superconductor}
\label{sect:reviewbcs}
Before moving to the holographic study of the single impurity effect in superconductor,  we will first briefly review the
results obtained in BCS theory with an impurity in the center of a superconductor in condensed matter physics.
 The reduced mean field BCS Hamiltonian of a pure superconductor can be written in momentum space as
\begin{equation}
H=\sum_{k} \varepsilon_k(c_{k \uparrow}^+c_{k \uparrow}+c_{-k \downarrow}^+c_{-k \downarrow})- \Delta_0 \sum_{k}(c_{k \uparrow}^+ c_{-k \downarrow}^+ +c_{-k \downarrow} c_{k \uparrow})+ \Delta_0^2/V,
\end{equation}
in which $\varepsilon_k= E_k- E_F$ with $E_F$ the fermi energy, $\Delta_0= V \sum_k<c_{-k\downarrow} c_{k\uparrow}>$ is the order parameter, and $V$ is the attractive interaction of the cooper pairs, which has a positive value due to the negative sign of the second term. One should note that $V$ is  non-zero close to the Fermi surface only when $|\varepsilon_k| < \hbar \omega_D$. The self-consistent
gap equation of $\Delta_0$ reads
\begin{equation}
1=V g(0) \int_0^{\hbar \omega_D}\frac{d \varepsilon}{\sqrt{\varepsilon^2+ \Delta_0^2}}.
\end{equation}
Where $g(0)$ is the state density at the Fermi surface. When a small impurity is added, it is
reasonable to assume that the state density $g(0)$ keeps the same as the pure case, thus the gap
$\Delta_1$ with an normal impurity still keeps the same as $\Delta_0$. This is exactly the Anderson theorem
explained for the early experiments that a superconductor is robust to a small normal impurity. \cite{anderson}

\begin{figure}[t]
\begin{center}
\includegraphics[trim=0cm 0cm 0cm 0cm, clip=true,scale=0.75]{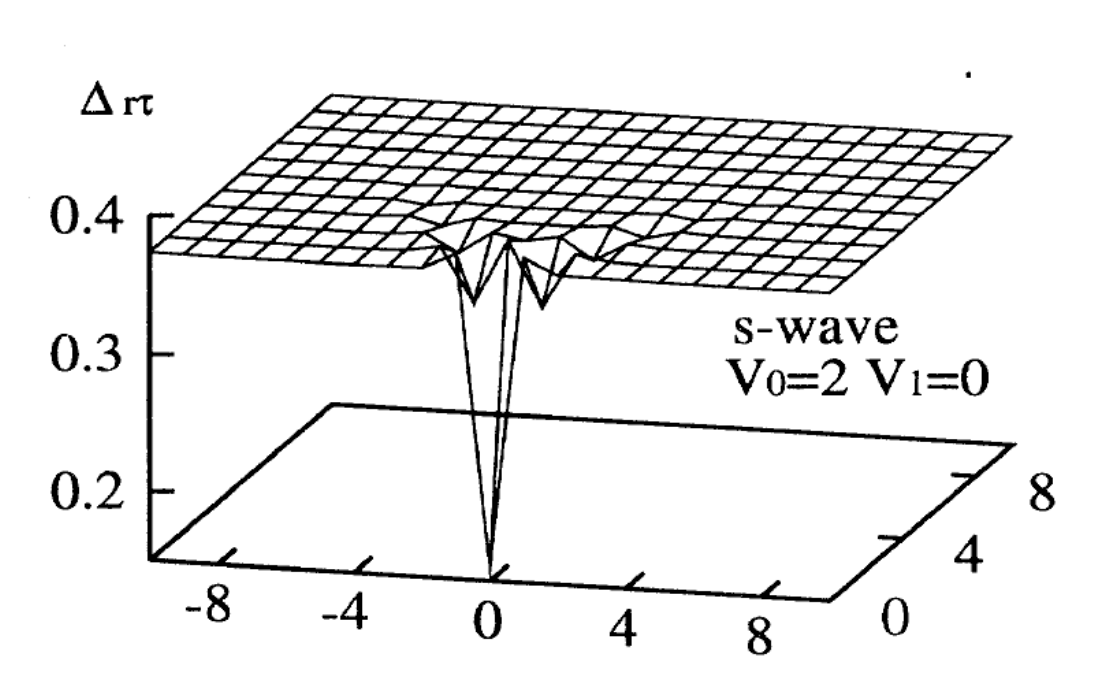}
\caption{ The 3d plot of the gap taken in \cite{xiang}, it shows the self-consistent gap function for the s-wave superconducting
state on a $21\times21$ lattice. Only half of the lattice
is shown. The impurity is located at the center of the lattice. Since
the scattering potential is short ranged (small impurity), the gap function
changes only in the vicinity of the impurity.}\label{fig1}
\end{center}
\end{figure}

We have to say that the Anderson theorem is an approximate statement, since the gap equation with an impurity scattering
is not solved exactly. In order to get an exact configuration of the gap in the presence of a normal impurity, the self-consistent BdG technique
is needed. Here we mainly review the results in \cite{xiang}, when an impurity is presented, we adopt the real space
Hamiltonian in square lattice as,
\begin{eqnarray}
H'[\Delta_{r\tau}]&=&-t \sum_{<r r'>}\left(c_{r \uparrow}^{+}c_{r'\uparrow}-c_{r\downarrow}^+ c_{r'\downarrow}\right)+\sum_{r\tau}\left(\Delta_{r\tau}c_{r\uparrow}^+c_{r+\tau \downarrow}^++H.c.\right)\nonumber\\
&&+\sum_{r}\left(\sum_{r_i}V_{r_i,r}-\mu\right)\left(c_{r \uparrow}^+c_{r \uparrow}-c_{r \downarrow}^+c_{r \downarrow}\right)+const.
\end{eqnarray}
In which $<rr'>$ indicates $r$ and $r'$ are the nearest neighbors, while $\mu$ is the chemical potential.
The effect of the impurity at $r_i$ is captured by the induced scattering potential $V_{r_i,r}$, for a single impurity $V_{r_i,r}=V_0 \delta_{r_i,r}$. $\Delta_{r\tau}$, which is independent of $r$, is the gap without
an impurity. However, in the presence of an impurity the gap
is $r$ dependent, and it can be obtained by solving the BdG equation numerically. The main
results in \cite{xiang} is plotted in Fig.\ref{fig1}, when the scattering is small  ($V_0=2$). The gap
at the impurity will be suppressed a lot, while the gap outside the impurity takes the same
value as the pure superconductor. In the paper \cite{xiang},
by increasing the impurity  to the strong scattering region ($V_0=20$), the author also found a decreasing gap of
the host superconductor, which in reverse indicates a decrease of critical temperature when a strong impurity is presented.

In summary, when the single impurity scattering is small, the host superconductor will hardly be affected by the
impurity; however, for large impurity scattering, the gap of the host superconductor will be reduced. There exists  a critical
impurity strength above which the gap will disappear. In the following sections, we will use the holographic method to study the
normal impurity effect to the superconductor, in which both the phenomena with weak and strong impurity scattering are similar to that in condensed matter physics.

\section{The Holographic Set-up}
\label{sect:anderson}

Even if the impurity concentration is small, the potential scattering induced by a local or finite size normal impurity in a homogeneous superconductor
will modify the properties (for example the gap and the charge density) of the host superconductor at the impurity point as reviewed in the above section. Other self-consistently determined non-uniform gap functions had also already been obtained in \cite{gapconfig1,gapconfig2} by solving the spatially dependent gap equation with a local impurity, from which the gap was strongly suppressed and localized in a small region.  For simplicity, we can consider the impurity effect by coupling a superconducting host to a small normal impurity in its center similar to \cite{xiang}.
From the gravity side,
the equations of motions (EoMs) of the scalar field and gauge field in the bulk correspond to
 the gap equations in the BCS theory.\cite{h1}
Moreover, from the AdS/CFT dictionary, chemical potential and charge density of the boundary field theory are dual to the coefficients of the expansions of the gauge field $A_t$ near the boundary, i.e., $A_t(z\to0, r)=\mu(r)-\rho(r) z$, in
which $z$ is the bulk radial coordinate while $r$ is the polar radial coordinate of the boundary spacetime.
In particular, we introduce a finite size impurity  in the center of the host by imposing
a boundary condition that at the center of the host (with a small finite
size), $\mu(r)$ (or $\rho(r)$) takes a smaller value, while outside the impurity point they take a larger value above the critical point. Thus
the host is in the superconducting phase. Our method to include a localized impurity in holographic superconductor
 is somewhat different from \cite{holoimpurity1,holoimpurity2},  in which the average effect of impurity
is studied holographically by introducing another massive gauge field,  this massive gauge field is supposed to dual to the added impurity.

Concretely, we adopt the action in the bulk which is dual to a holographic superconductor \cite{h1} as,
\begin{equation}\label{action}
S=\int d^4x\sqrt{-g}[R-2\Lambda-\frac14F_{\mu\nu}F^{\mu\nu}-|\nabla\psi-iA\psi|^2-m^2|\psi|^2],
\end{equation}
where $\Lambda=-d(d-1)/2\ell^2$ is the cosmological constant, $d$ is the dimension of the boundary, $\ell$ is the radius of the AdS spacetime, and $F_{\mu\nu}=\partial_\mu A_\nu-\partial_\nu A_\mu$ is the strength of the gauge field. The metric is an AdS-Schwarzschild black hole,
\begin{equation}
{ds^2=\frac{\ell^2}{z^2}(-h(z)dt^2+ dr^2+r^2 d\theta^2)+\frac{\ell^2 dz^2}{z^2 h(z)},}
\end{equation}
{with $h(z)=1-z^3/z_0^3$, where $z_0$ is the position of horizon.}  In which $r,\theta$ are the boundary radial and angle coordinates respectively(
we use polar coordinates on the boundary in order to put an impurity at the center of the host)
. Without loss of generality, we set $ \ell=1$.
The temperature of the black hole is $T=\frac{3 }{4 \pi z_0}$, { besides we set $z_0=1$ in the following context}.
We use the ansatz that $\psi=\psi(z,r)$, $A=(A_t(z,r),0,0,0)$,
and $m^2=-2$.
In the probe limit, with the rescaling of $\psi\rightarrow \psi z$, we have the following EoMs:
\begin{eqnarray}
\left(1-z^3\right) \partial_z^2A_t+\partial_r^2 A_t + \frac{1}{r} \partial_r A_t-2
    A_t  \psi ^2=0,&&\label{eqnAt}\\
   \psi  \left( A_t^2+z^4-z\right)+\left(1-z^3\right)
   \partial_r^2\psi + \frac{1}{r} \left(1-z^3\right) \partial_r \psi +\left(z^3-1\right)^2 \partial_z^2\psi  +3
   \left(z^3-1\right) z^2 \partial_z\psi =0.&&\label{eqnpsi}
\end{eqnarray}
The expansions of $\psi$ and $A_t$ near the infinite boundary are:
\begin{eqnarray}\label{eomr}
\psi(z,r)&\sim&\psi^{(0)}(r)+\psi^{(1)}(r) z+\dots,\\
A_t(z,r)&\sim&\mu(r)-\rho(r) z+\dots.
\end{eqnarray}
From the AdS/CFT dictionary, $\psi^{(0)}$ is interpreted as the source of the boundary scalar operator while $\psi^{(1)}$ can be regarded as the condensate value of the operator. In the holographic superconductors, we usually turn off the source of the scalar operator, i.e., $\psi^{(0)}=0$ since
we expect a spontanous symmetry broken of the boundary theory.
It has been confirmed that, in the homogeneous case, there is a continuous phase transition from the normal state (perfect metal state) to the
superconducting state with the usual mean-field critical exponent $1/2$ by reducing the temperature\cite{h1}.
The critical temperature $T_{c0}$ of the phase transition is $T_{c0}\approx 0.0588\mu$ in unit of chemical potential.
In the paper we also plot all the dimensionless quantities in  the unit of chemical potential.
In order to simulate the single normal impurity effect in the center of the sample,  we introduce a chemical potential in the polar coordinates  as
\begin{equation}
\mu(r)=\mu_{max}\left\{1-\frac{\epsilon}{2\tanh(\frac{L}{2\sigma})}
  \left[1
  -\tanh\left(\frac{r-\tfrac{L}{2}}{\sigma}\right)\right]\right\}\;,
\end{equation}
where $\mu_{max}$ is the chemical potential outside the impurity, and
the parameters $L/2$, $\sigma$ and $\epsilon$ are the radius,  steepness and depth of the impurity respectively. The maximal value of  $\epsilon$ is $1$ in order to insure $\mu(r)$ is always positive. We can also introduce a charge density with similar form to get similar results.
We emphasize that the exact form of $\mu(r)$ or $\rho(r)$ is not important.
At the largest $r=r_{max}$,  where $r_{max}$ is the size of the host superconductor we choose,
we used  the boundary condition that $\partial_r At=\partial_r \psi=0$.
This is physical since the $\mu(r)$ is independent of $r$ and the condensate is also independent of $r$. This is important when we derive the
expression of the free energy in Appendix A.

\begin{figure}[t]
\begin{center}
\includegraphics[trim=2cm 9.5cm 2cm 9.5cm, clip=true,scale=0.55]{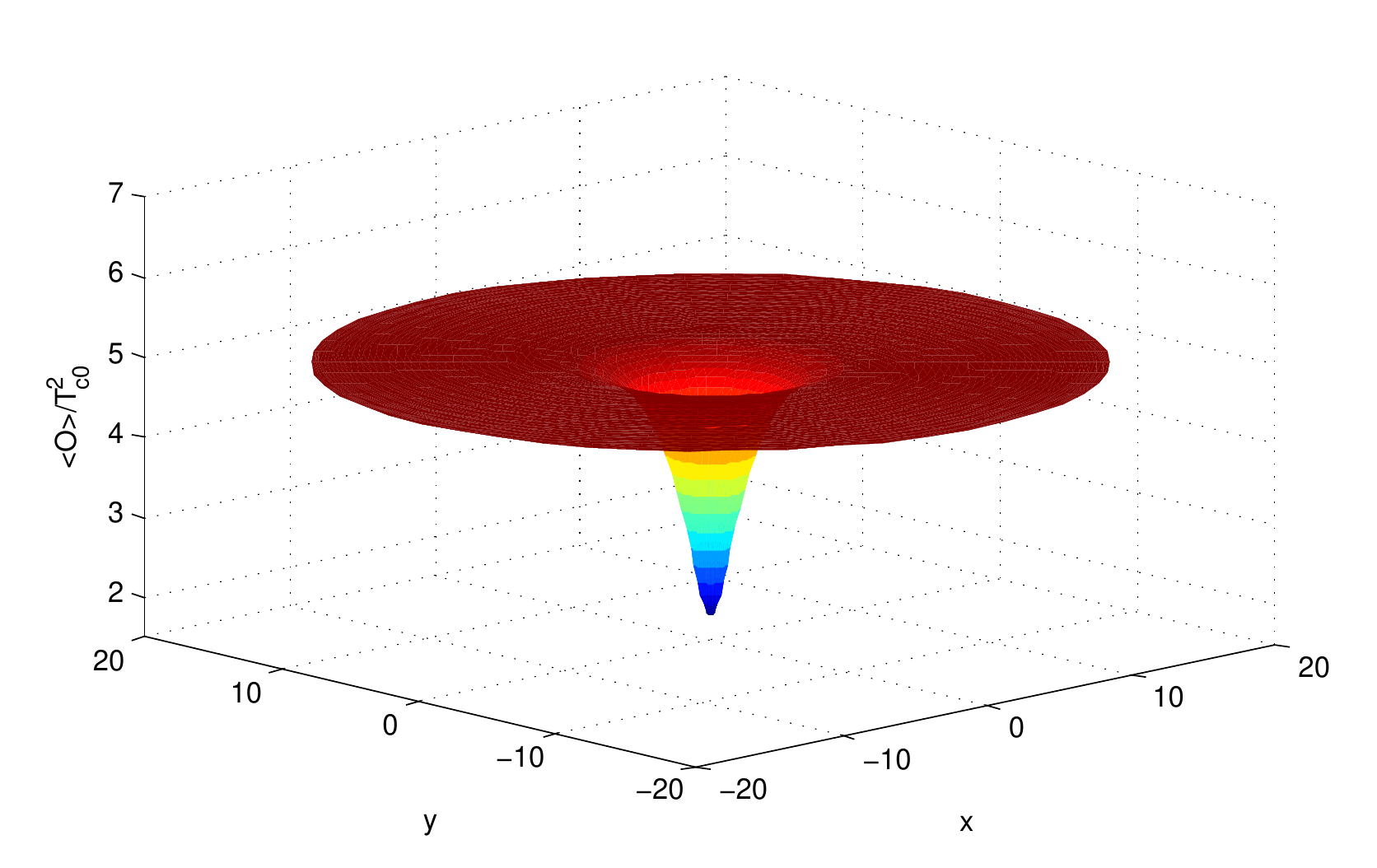}
\caption{A 3d plot of order parameter with $\mu_{max}=4.2$, $L=2$, $\sigma=0.5$ and $\epsilon=0.2$.  The size of the host is set to be $r_{max}=20$, while the impurity is located in the center of the host with a small finite size.}\label{fig2}
\end{center}
\end{figure}

\begin{figure}[h]
\begin{center}
\includegraphics[trim=1cm 9.5cm 2cm 13.5cm, clip=true,scale=0.85]{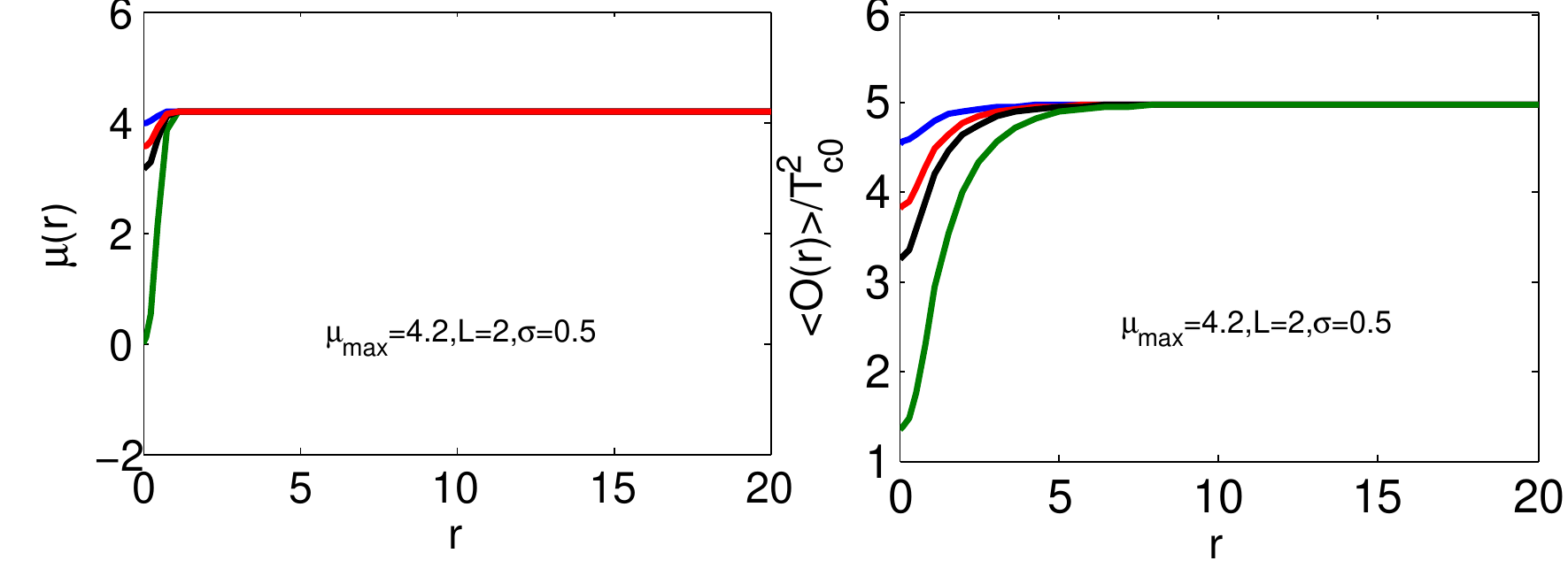}
\caption{The configuration of $\mu(r)$ (left) and the condensate of the order parameter (right) for different depths of the impurity.
The parameters are $\mu_{max}=4.2$, $L=2$, $\sigma=0.5$ and four different $\epsilon=0.05;0.15;0.25;1$ (from top to bottom).
The size of the host $r_{max}$ is fixed to be 20.}\label{fig3}
\end{center}
\end{figure}

The EoMs Eq.\eqref{eqnAt} and Eq.\eqref{eqnpsi}  are solved by the the Chebyshev spectral methods \cite{spectral}.
We discretize the system on a two dimensional Chebyshev grids with $20$ points along the
$z$ direction while $80$ points in the $r$ direction. A sample plot of the order parameter configuration is shown in Fig.\ref{fig2} with $\mu_{max}=4.2$, $\sigma=0.5$ and $\epsilon=0.2$, while the size (radius) of the host is $r_{max}=20$ and the impurity size is $L/2=1$.
It is clear that at the impurity point (the center of the host) the gap is suppressed a lot compared to the host superconductor.
This gap configuration is
very similar to the results obtained in \cite{xiang} (Fig.\ref{fig1} therein), as well as Fig.2 in \cite{gapconfig2}, in which the inhomogeneous
gap was obtained by solving the self-consistent BdG equation with an impurity at the center.
In order to see that the host superconductor will not be affected by an impurity of any depth, we fixe the radius $L/2=1$ and the host size $r_{max}=20$.
The gap configuration for any depth $\epsilon$ can be found on the right panel of Fig.\ref{fig3}, in which we can see that around $r>9$ the order parameters have the same values whatever the depths are.  This also indicates that the critical temperature $T_c$  for the host superconductor dose not change with respect to the depth, as Anderson theorem stated.

\section{Suppression of Superconductivity}
\label{sect:suppresion}

\begin{figure}[t]
\begin{center}
\includegraphics[trim=3cm 9.5cm 5cm 12.cm, clip=true,scale=0.65]{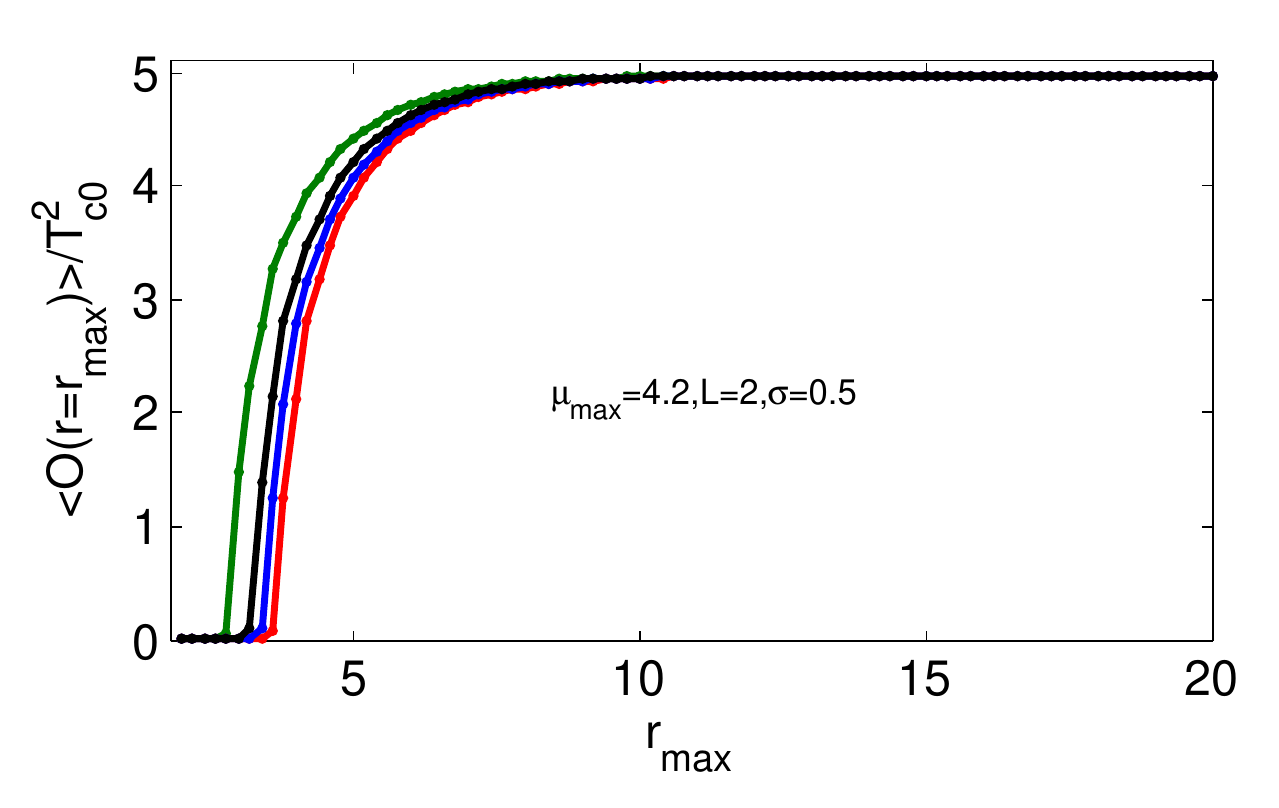}
\caption{The condensate at  $r=r_{max}$ for different sizes of the host superconductor $r_{max}$ with fixed $\mu_{max}=4.2$, and fixed impurity
 size  $L/2=1$. The four lines from top to bottom correspond to $\epsilon=0.3;0.5;0.7;1$.
 }\label{fig4}
\end{center}
\end{figure}

   Since the order parameter is reduced a lot at the impurity point, it is natural to
expect that by increasing the impurity size with fixed  host superconductor size,
or by reducing the host size while fixing the impurity size, the order parameter
of the whole host superconductor will be suppressed. This phenomenon is shown in Fig.\ref{fig4},
in which we plot the condensate at $r_{max}$ for different host superconductor
size $r_{max}$ with fixed impurity size $L/2=1$ and $\mu_{max}=4.2$. We can see that
for different $\epsilon = 0.3; 0.5; 0.7; 1$, there is a critical $r_{max}^{c1} \simeq 8$,
below which the condensate of the  host superconductor will become suppressed.
     Further more, for different depth of the impurity we
     see different values of critical $r_{max}^{c2}$ below which the order parameter
     vanish. Larger $\epsilon$ have a larger value of $r_{max}^{c2}$, which is reasonable
     since for larger depth of impurity the order parameter is more easy to be destroyed.

\begin{figure}[h!]
\begin{center}
\includegraphics[trim=0cm 10cm 2cm 11cm, clip=true,scale=0.65]{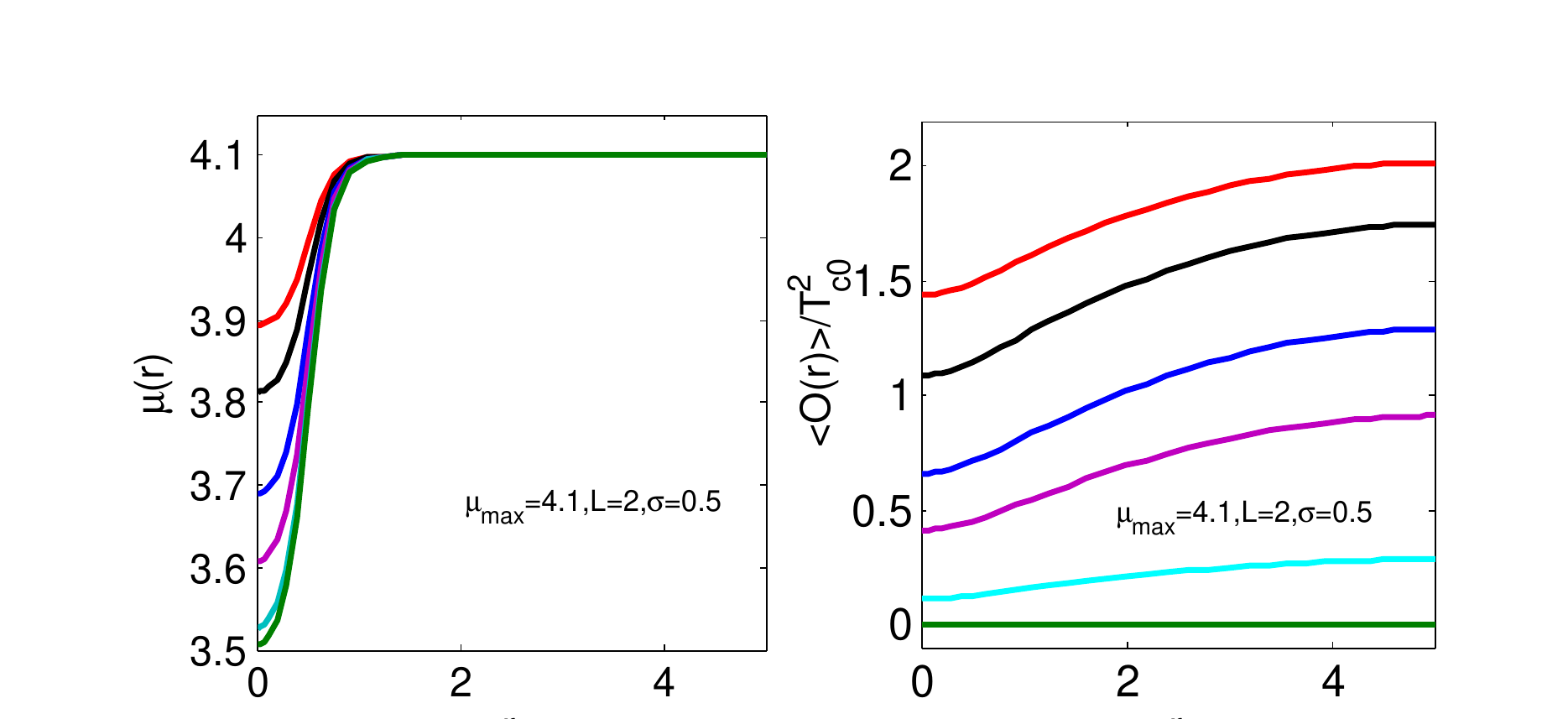}
\caption{The configuration of $\mu(r)$ (left) and the condensate of the order parameter (right) for different depths of the impurity with small host superconductor size $r_{max}=5$.
We choose $\mu_{max}=4.1$, $L=2$,$\sigma=0.5$ and six different $\epsilon=0.05;0.07;0.1;0.12;0.14;0.145$ from top to bottom.}\label{fig5}
\end{center}
\end{figure}

\subsection{The critical depth of impurity $\epsilon_c$}
With the realizations above, we take $r_{max}=5$  and fix the impurity size $L/2=1$ as an example.
We find that increasing the strength of the impurity (associated to the depth $\epsilon$) will finally induce a phase transition from superconducting
state to normal state, see Fig.\ref{fig5}. From Fig.\ref{fig5} we can see that for a host superconductor with temperature  $T \propto 1/\mu_{max}$, where $\mu_{max}=4.1$, which is close to $T_{c0} \propto 1/\mu_{c0}$ with $ \mu_{c0}=4.06$,
the increasing impurity depth will suppress the host superconductor and finally destroy the superconducting condensate.
The phase transition occurs at about $\epsilon \simeq 0.15$.

\begin{figure}[t]
\begin{center}
\includegraphics[trim=0cm 8.5cm 6cm 10.5cm, clip=true,scale=0.65]{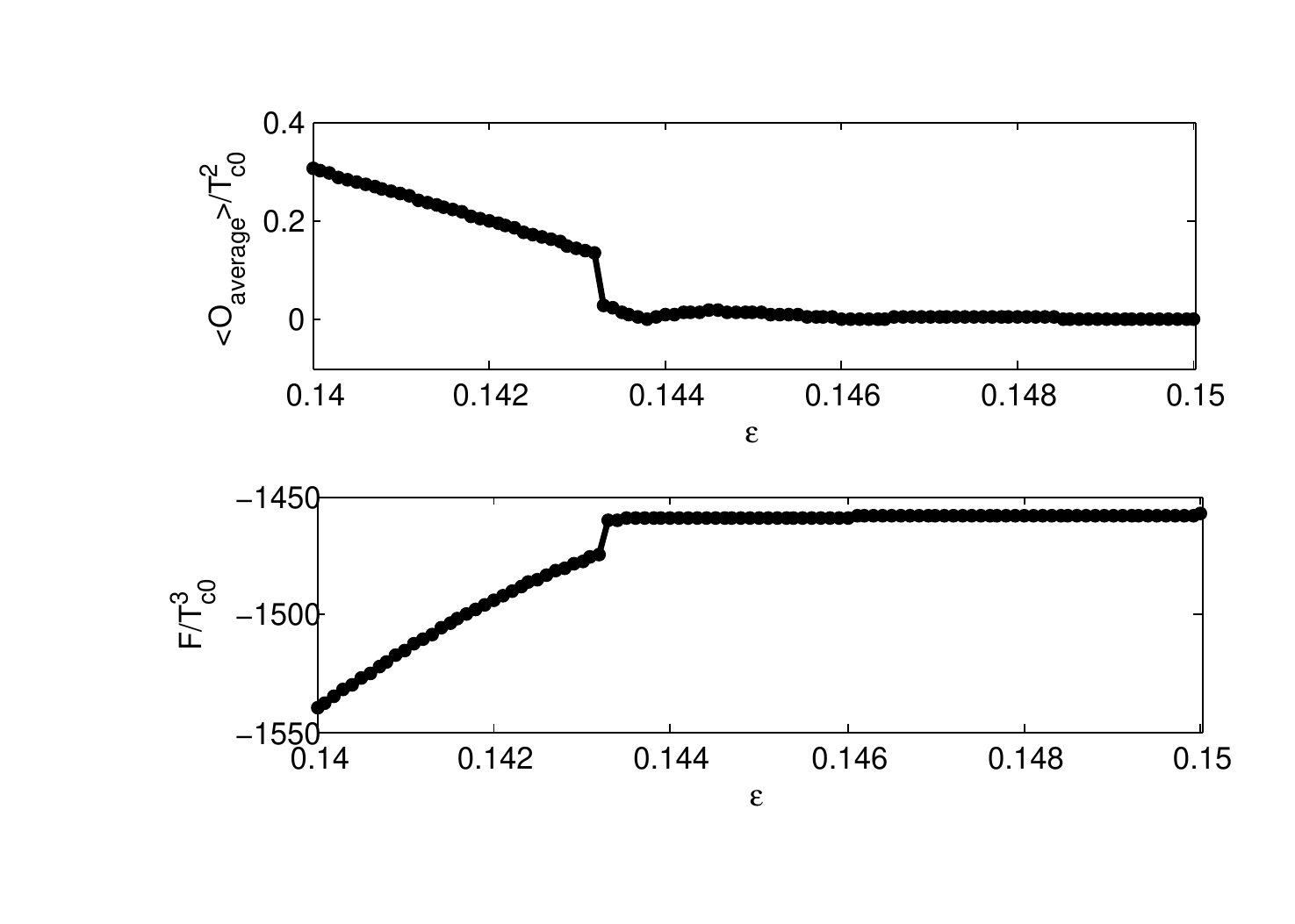}
\caption{The phase transition driven by the increase of impurity depth $\epsilon$. The parameters are $\mu_{max}=4.1$, $L=2$,$\sigma=0.5$ and $r_{max}=5$.
The discontinuous property of the free energy(bottom) at the critical point indicates that the phase transition is of zeroth order. }\label{fig6}
\end{center}
\end{figure}

\begin{figure}[h]
\begin{center}
\includegraphics[trim=0cm 9.5cm 6cm 10.5cm, clip=true,scale=0.65]{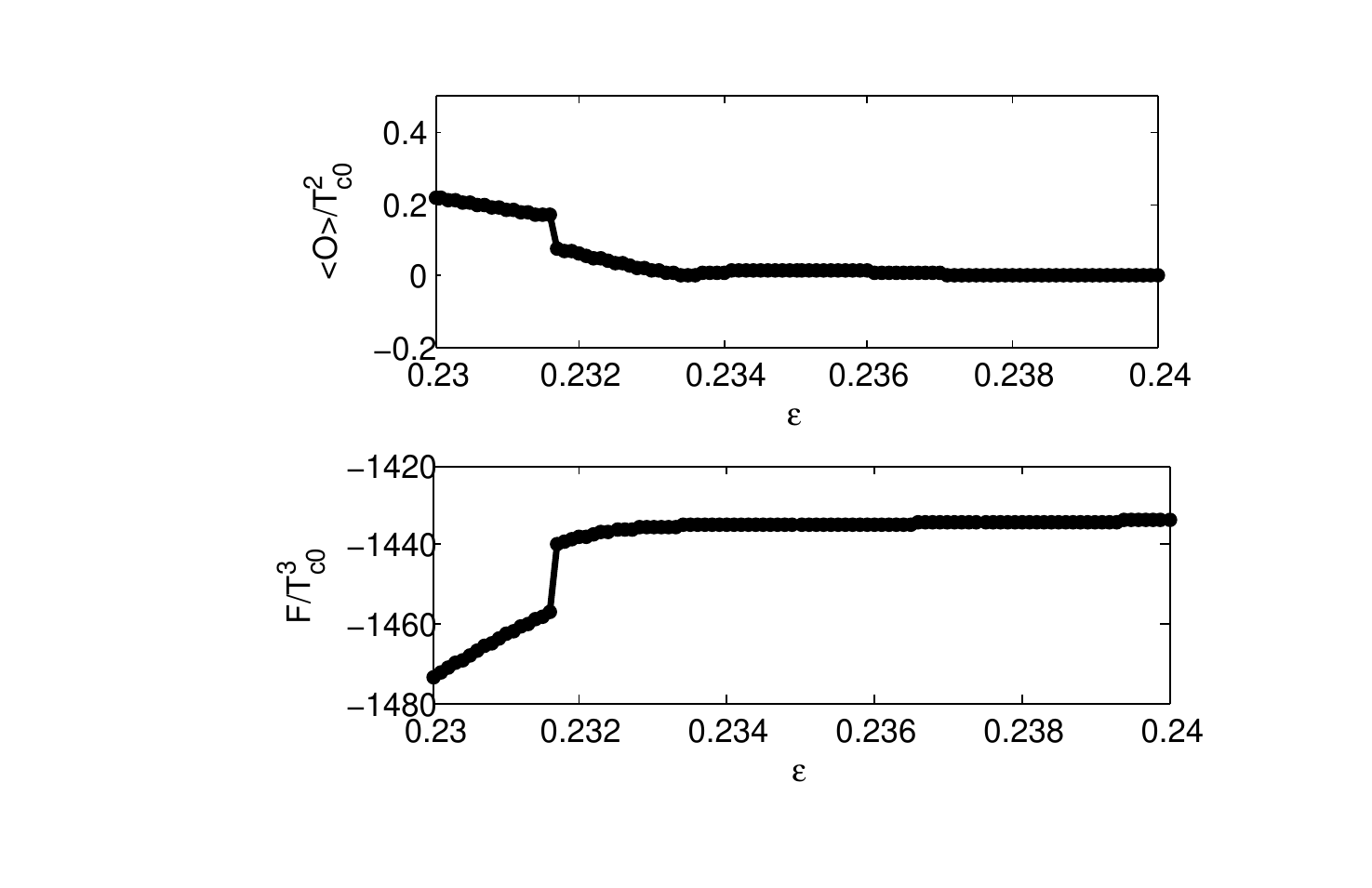}
\caption{The phase transition driven by the increase of impurity depth $\epsilon$. The parameters are $\mu_{max}=4.11$, $L=2$, $\sigma=0.5$ and $r_{max}=5$.
The discontinuous property of  free energy(bottom) at the critical point indicates that the phase transition is of zeroth order.}\label{fig7}
\end{center}
\end{figure}

\begin{figure}[h!]
\begin{center}
\includegraphics[trim=0cm 0cm 0cm 0cm, clip=true,scale=0.85]{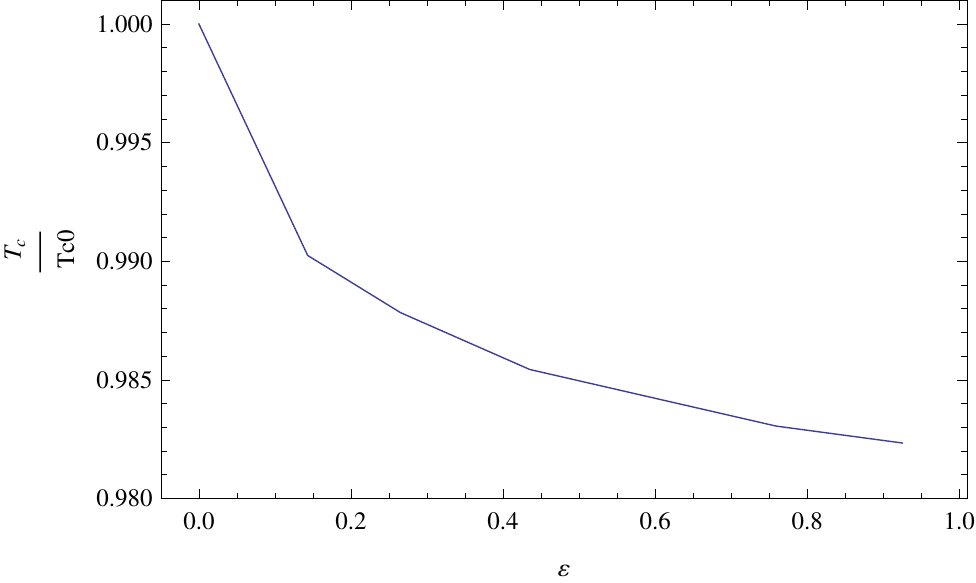}
\caption{Critical temperature decreases when increasing the impurity depth with
fixed $L=2$, $\sigma=0.5$ and $r_{max}=5$. }\label{fig8}
\end{center}
\end{figure}

In order to find the exact value of $\epsilon_c$ where the superconductor/metal phase transition
occurs, we scanned $100$ points from $\epsilon=0.14$ to $\epsilon=0.15$ with every step as $10^{-3}$.
The results are shown in Fig.\ref{fig6}, we can see that $\epsilon_c\sim0.143$ when  $\mu_{max}=4.1$.
Another case is shown in Fig.\ref{fig7}, with a larger $\mu_{max}=4.11$, the critical depth of impurity is $\epsilon_c\approx0.232$.
The phase diagram with $L=2$, $\sigma=0.5$ and fixed $r_{max}=5$ is plotted in Fig.\ref{fig8}. The critical temperature decreases
with increasing $\epsilon$. Though the reduction is small, we can still see a phase transition when the host superconductor
condensate is small.

The discontinuous order parameter shown in Fig.\ref{fig6} and Fig.\ref{fig7} indicate that the phase transition at the critical $\epsilon_c$ is also discontinuous.
In order to prove the order of phase transition we need to compute the free energy.


\subsection{Discontinuous free energy at the phase transition point }
To find the order of phase transition we need to compute the free energy. According to the AdS/CFT dictionary, the free energy of the boundary
theory is given by the on-shell action of the bulk theory, $F=-TS_{o.s.}$ \cite{Skenderis:2002wp}.
In the holographic superconductor model,  $S_{o.s.}$ can be evaluated from integrating by parts and using the equations of motion as well as suitable counter terms, which results in\cite{herzog},
\begin{equation}
F \propto \int_{z=0} d^3 x (\frac{A_t \partial_z A_t}{2})+ \int d^4 x \frac{\psi^2 A_t^2}{(1-z^3) z^2},
\end{equation}
in which $\psi$ is from the  action \eqref{action}.
The details of deriving the expression  of the free energy can be found in the Appendix \ref{sect:appendix}.
The results shown in Fig.\ref{fig6} and Fig.\ref{fig7} tell us  that at the critical point where the phase transition occurs,
the free energy is also discontinuous, which confirms that the phase transition is of zeroth order.
The  finding of zeroth phase transition is somehow unexpected but interesting.
Even in the condensed matter literatures \cite{BVZ}, people did not study the order of
the  phase transition, then what we predict here can be potentially confirmed by experiments.
Here we fix the temperature while increasing the impurity depth to trigger the phase transition,
which is equivalent to fix an impurity strength  while increasing the temperature which is lower than $T_c$ to the same
critical point in Fig.\ref{fig8}.
Other observations of zeroth order phase transition  can be found in \cite{p7,lili} in holographic superconductor and
also in a rotating AdS black hole\cite{robert}.
 In particular, in \cite{p7} the zeroth order phase transition is also triggered by an inhomogeneous source on the boundary.
What we found  here is indeed similar to \cite{p7} since in both cases the transitions are triggered by
inhomogeneous charge density of the boundary field theory. However, in \cite{lili} the authors found that
if redone considered the strong back-reaction of the condensed fields and gauge fields in a homogeneous $p$-wave
holographic superconductor, by reducing temperature  one would see the zeroth order phase transition.
This is different from what we found since we are  working in the probe limit,  and also the zeroth phase transition
is triggered by the inhomogeneous charge density. In \cite{robert} the zeroth order
phase transition was found in a $d=6$ dimensional rotating black hole by
reducing temperature, this is also different from what we found.
Although in these interesting works the zeroth order phase transition has been
found, it is hard to uncover the common  features behind these different models.

\section{Conclusions and Discussions}
\label{sect:conclusion}

In this paper, the single normal impurity effect is investigated
in the holographic $s$-wave superconductor. We uncover that the host superconductor is robust to the
small size impurity, as the Anderson theorem stated \cite{anderson,ma}.
However, although the host superconductor is robust to the impurity,
the gap at the impurity site is strongly suppressed, which
agrees with the studies by solving the BdG equation in the presence of
a normal impurity \cite{gapconfig1,gapconfig2} in condensed matter physics. The suppression of the gap at the impurity point indicates
a decrease of host order parameter if we have a larger size
impurity while fixing the host size, or equivalently a decrease of the host superconductor
size while fixing the impurity size.
This is similar to the studies in superconductors with ultrashort coherence length and a superconductor
with strong impurity scattering, in which
the host order parameter will be reduced substantially when an impurity is presented \cite{ghosal,xiang}.
Moreover, if we have a small  host superconductor
with small condensate we obtain a phase transition from superconducting
state to normal metal state by increasing the impurity depth.  These phenomena
are similar to the results in the condensed matter literatures about the normal
impurity effect in a superconductor\cite{BVZ}. The impurity induced phase transition
is of zeroth order which might be observed in future experiments. A promising suggestion was that the entanglement entropy may be a good observable to study the zeroth phase transition in our model. 

As another important topic in condensed matter physics, to understand the
magnetic impurity effect in a superconductor holographically is also important, a possible way to study the
magnetic impurity effect in holographic superconductor is to adopt the holographic paramagnetism-ferromagnetism phase transition\cite{cai} in a small region of a
large host superconductor. We expect that by coupling a holographic superconductor to a magnetic impurity, the
Kondo effect \cite{kondo} can be realized holographically as in \cite{holokondo1,holokondo2}.

\textbf{Acknowledgments:} We thank Yong-Qiang Wang, Prof. Chiang-Mei Chen, Prof. Rong-Gen Cai, Li Li and  Phil Szepietowski for many valuable comments. The authors are  grateful to the Mainz Institute for Theoretical Physics (MITP) and CERN for their hospitality and their partial support during the completion of this work. This work is supported in part by the National Natural Science Foundation of China (Grant No. 11205020 and No.11205097) and in part by the fund of Utrecht University budget associated to Gerard 't Hooft.

\begin{appendix}
\section{Derivation of the free energy $F$}
\label{sect:appendix}
The generic on-shell action for the Einstein-Maxwell-scalar action \eqref{action} (in the probe limit) is \cite{Garcia-Garcia:2013rha}
\begin{eqnarray}
S_{\rm{o.s.}}=-\frac12\int d^4x \partial_\mu\left[\sqrt{-g}\left(A_\nu F^{\mu\nu}+\psi^*\partial^\mu\psi+\psi\partial^\mu\psi^*\right)\right]&&\nonumber\\
+\frac{iq}{2}\int d^4x\sqrt{-g}A_\mu\left(\psi^*\partial^\mu\psi-\psi\partial^\mu\psi^*-2iqA^\mu\psi\psi^*\right).&&
\end{eqnarray}
The first integral of $S_{\rm{o.s.}}$ is a surface term, which can be further reduced, according to the ansatz in this paper, as
\begin{eqnarray}\label{surf2}
S_{\rm{surf.}}&=&-\frac12\int d^4x \partial_\mu\left[\sqrt{-g}\left(A_\nu F^{\mu\nu}+\psi^*\partial^\mu\psi+\psi\partial^\mu\psi^*\right)\right]\nonumber\\
&=&-\frac12\int dtdrd\theta~ r\left(-A_t\partial_zA_t+\frac{2h}{z^2}\psi\partial_z\psi\right)\bigg|^{z=1}_{z=0}-\frac12\int dtdzd\theta ~r \left(-\frac1h A_t\partial_rA_t+\frac{2}{z^2}\psi\partial_r\psi\right)\bigg|^{r_{\rm{max}}}_{r=0}\nonumber\\
\end{eqnarray}
The last term in the above eq.\eqref{surf2} is vanishing since we have imposed the flat boundary condition of the fields at $r=r_{\rm{max}}$, i.e., $\partial_rA_t(r_{\rm{max}})=\partial_r\psi(r_{\rm{max}})\equiv0$. Therefore, the surface term now becomes,
\begin{eqnarray}
S_{\rm{surf.}}&=&-\frac12\int dtdrd\theta~ r\left(-A_t\partial_zA_t+\frac{2h}{z^2}\psi\partial_z\psi\right)\bigg|^{z=1}_{z=0}=\frac12\int dtdrd\theta~ r\left(-A_t\partial_zA_t+\frac{2h}{z^2}\psi\partial_z\psi\right)\bigg|_{z=0}\nonumber\\
&=&\frac12\int dtdrd\theta~ r\left(-A_t\partial_zA_t+\frac{2}{z^2}\psi\partial_z\psi\right)\bigg|_{z=0}.
\end{eqnarray}
in which, we have adopted $A_t(z=1)=h(z=1)=0$ and $h(z=0)=1$. In fact, the surface term above is divergent at $z=0$, we need to add a counter term into the on-shell action, which is
\begin{eqnarray}
S_{\rm{ct.}}=-\int dtdrd\theta\sqrt{-\gamma}\psi^2\bigg|_{z=0}
\end{eqnarray}
where $\gamma$ is the reduced metric on the cutoff surface $z=0$, and $\sqrt{-\gamma}=r/z^3$. Therefore, the finite surface term is
\begin{eqnarray}
S_{\rm{fi.}}=S_{\rm{surf.}}+S_{\rm{ct.}}=\int dtdrd\theta~r\left(-\frac12A_t\partial_zA_t+\psi^{(0)}\psi^{(1)}\right)\bigg|_{z=0}=\int dtdrd\theta~r\left(-\frac12A_t\partial_zA_t\right)\bigg|_{z=0}.
\end{eqnarray}
We have used $\psi^{(0)}(z=0)=0$ in the last step. Therefore, the total finite on-shell action is,
\begin{eqnarray}
S_{\rm{o.s.}}&=&S_{\rm{fi.}}+\frac{iq}{2}\int d^4x\sqrt{-g}A_\mu\left(\psi^*\partial^\mu\psi-\psi\partial^\mu\psi^*-2iqA^\mu\psi\psi^*\right)\nonumber\\
&=&\int dtdrd\theta~r\left(-\frac12A_t\partial_zA_t\right)\bigg|_{z=0}-q^2\int dtdzdrd\theta\frac{r}{z^2h}A_t^2\psi^2\nonumber\\
&=&-\int_{z=0}dtdx_1dx_2\left(\frac{A_t\partial_zA_t}{2}\right)-q^2\int dtdzdx_1dx_2\left(\frac{A_t^2\psi^2}{z^2(1-z^3)}\right)
\end{eqnarray}
In the last step we have recovered the polar coordinates $(r,\theta)$ on boundary to the usual Cartesian coordinates $(x_1, x_2)$ {\blue with $x_1=r\cos\theta$ and $x_2=r\sin\theta$}. Therefore, the free energy $F$ is
\begin{eqnarray}
F=-TS_{\rm{o.s.}}\propto  \int_{z=0} d^3 x \left(\frac{A_t \partial_z A_t}{2}\right)+ q^2\int d^4 x \frac{\psi^2 A_t^2}{(1-z^3) z^2}.
\end{eqnarray}

\end{appendix}


\begin{thebibliography}{99}

 \bibitem{adscft}
  J.~M.~Maldacena,
  Adv.\ Theor.\ Math.\ Phys.\  {\bf 2} (1998) 231;
  S.~S.~Gubser, et al.,
  Phys.\ Lett.\  B {\bf 428} (1998) 105;
  E.~Witten,
  Adv.\ Theor.\ Math.\ Phys.\  {\bf 2} (1998) 253.

\bibitem{r1} S. A. Hartnoll, ``Lectures on holographic methods for condensed matter physics'',  Class.Quant.Grav. 26 (2009) 224002, [arXiv:0903.3246].
\bibitem{r2} C. P. Herzog, ``Lectures on Holographic Superfuidity and Superconductivity'', J.Phys.  A42 (2009) 343001, [arXiv:0904.1975].
\bibitem{r3} J. McGreevy, ``Holographic duality with a view toward many-body physics'', Adv.High  Energy Phys. 2010 (2010) 723105, [arXiv:0909.0518].
\bibitem{r4} S. Sachdev, ``Condensed Matter and AdS/CFT'', Lect.Notes Phys. 828 (2011) 273-311,  [arXiv:1002.2947].
\bibitem{h1}S.~S.~Gubser,``Breaking an Abelian gauge symmetry near a black hole horizon'', Phys.\ Rev.\  D {\bf 78} (2008) 065034, arXiv:0801.2977 [hep-th]; S.~A.~Hartnoll, C.~P.~Herzog and G.~T.~Horowitz, ``Building an AdS/CFT superconductor'', Phys.\ Rev.\ Lett.\  {\bf 101} (2008) 031601, arXiv:0803.3295 [hep-th].

\bibitem{r5} G. T. Horowitz, "Introduction to Holographic Superconductors", [arXiv:1002.1722[hep-th]].

\bibitem{r6} D. Musso, "Introductory notes on holographic superconductors",[arXiv:1401.1504[hep-th]].

\bibitem{p0}
  A.~Donos and J.~P.~Gauntlett,
  ``Holographic striped phases,''
  JHEP {\bf 1108}, 140 (2011)
  [arXiv:1106.2004 [hep-th]].

\bibitem{p1} R. Flauger, E. Pajer and S. Papanikolaou, ``A Striped Holographic Superconductor'', Phys. Rev. D {\bf83}
 064009 (2011)[arXiv:1010.1775 [hep-th]].

\bibitem{p2} J. Erdmenger, X. -H. Ge and D. -W. Pang, ``Striped phases in the holographic insulator/superconductor transition'', JHEP 11 (2013) 027, arXiv:1307.4609[hep-th].

\bibitem{p3} S. Ganguli, J. A. Hutasoit, G. Siopsis,``Enhancement of Critical Temperature of a Striped Holographic Superconductor'',
 Phys. Rev. D {\bf86}, 125005 (2012).

\bibitem{p4} G. T. Horowitz and J. E. Santos, ``General Relativity and the Cuprates'',  [arXiv:1302.6586 [hep-th]].

\bibitem{p5} D. Arean, A Farahi, L. A. P. Zayas, I. S. Landea, A. Scardicchio,``A Dirty Holographic Superconductor'', Phys. Rev. D 89, 106003 (2014),	arXiv:1308.1920 [hep-th].

\bibitem{p6} X.-Mei Kuang, B. Wang, X.-H. Ge, ``Observing the inhomogeneity in the holographic models of superconductors'',  arXiv:1307.5932.

\bibitem{p7} H. B. Zeng, ``Anderson localization in a holographic superconductor'', Phys. Rev. D 88, 126004 (2013), arXiv:1310.5753 [hep-th].

\bibitem{p8} D. Arean, A. Farahi, L. A. Pando Zayas, I. S. Landea, and A. Scardicchio,  ``Holographic p-wave Superconductor with Disorder," arXiv:1407.7526  [hep-th].

\bibitem{p9} N.  Iizuka, K. Maeda, ``Towards the Lattice Effects on the Holographic Superconductor'', JHEP  1211 (2012) 117, [arXiv:1207.2943[hep-th]];

\bibitem{p10} F. Aprile and T. Ishii,`` A Simple Holographic Model of a Charged Lattice'', arXiv:1406.7193.

\bibitem{p11} Y. Ling, P. Liu, C. Niu, J. P. Wu, and Z. Y. Xian, Holographic Superconductor on Q-lattice, arXiv:1410.6761.

\bibitem{GarciaGarcia:2012zd}
  A.~M.~Garcia-Garcia, J.~E.~Santos and B.~Way,
  ``Holographic Description of Finite Size Effects in Strongly Coupled Superconductors,''
  Phys.\ Rev.\ B {\bf 86}, 064526 (2012)
  [arXiv:1204.4189 [hep-th]].

\bibitem{BVZ}
A. V. Balatsky, I. Vekhter, Jian-Xin Zhu, ``Impurity-induced states in conventional and unconventional superconductors '', Rev. Mod. Phys. 78, 373 (2006).

\bibitem{anderson}
P. W. Anderson, ``Theory of dirty superconductors", J. Phys. Chem. Solids 11, 26 (1959).

\bibitem{gapconfig1}
Franz, M., C. Kallin, and A. J. Berlinsky, ``Impurity scattering and localization in d-wave superconductors", Phys. Rev. B
54, R6897, (1996).


\bibitem{gapconfig2}
Tao Zhou, Xiang Hu, Jian-Xin Zhu, and C. S. Ting, ``Impurity effect as a probe for order parameter symmetry in iron-based superconductors'', arXiv:0904.4273.



\bibitem{ma} M. Ma and P. A. Lee, ``Localized superconductors", Phys. Rev. B 32, 5658 (1985).

\bibitem{ghosal} Ghosal, A., M. Randeria, and N. Trivedi, ``Role of Spatial Amplitude Fluctuations in Highly Disordered s-Wave Superconductors", Phys. Rev. Lett. 81, 3940 (1998).

\bibitem{xiang} Xiang, T., and J. M. Wheatley, ``Nonmagnetic impurities in two-dimensional superconductors", Phys. Rev. B 51, 11721 (1995).


\bibitem{holoimpurity1}  K. Hashimoto and N. Iizuka, ``Impurities in Holography and Transport Coefficients'', arXiv:1207.4643  [hep-th].

\bibitem{holoimpurity2} T. Ishii and S.-J. Sin, ``Impurity effect in a holographic superconductor'', JHEP 1304 (2013)  128, [arXiv:1211.1798].
\bibitem{spectral} L. N. Trefethen,{\it Spectral methods in MATLAB}, Siam, Philadelphia, (2000).

\bibitem{Skenderis:2002wp}
  K.~Skenderis,
  ``Lecture notes on holographic renormalization,''
  Class.\ Quant.\ Grav.\  {\bf 19}, 5849 (2002)
  [hep-th/0209067].

\bibitem{herzog} C. P. Herzog, P. K. Kovtun and D. T. Son, ``Holographic model of superfuidity,'' Phys.  Rev. D 79 (2009) 066002 [arXiv:0809.4870 [hep-th]].

\bibitem{lili} R. -G. Cai, L. Li, L. -F. Li, ``A Holographic P-wave Superconductor Model'', JHEP 1401,032 (2014) arXiv:1309.4877 [hep-th].
R. -G. Cai, L. Li, L. -F. Li and R. -Q. Yang, ``Towards Complete Phase Diagrams of a
Holographic P-wave Superconductor Model,'' JHEP 1404, 016 (2014) [arXiv:1401.3974 [gr-qc]].

\bibitem{robert} N. Altamirano, D. Kubiznak, and R.B. Mann,``Reentrant Phase Transitions in Rotating AdS Black Holes'', Phys. Rev. D 88, 101502 (2013) arXiv:1306.5756.

\bibitem{cai} R. -G. Cai and R. -Q. Yang, ¡°Paramagnetism-Ferromagnetism Phase Transition in a Dyonic  Black Hole,¡± arXiv:1404.2856 [hep-th]; R. -G. Cai and R. -Q. Yang, ¡°A Holographic Model for Paramagnetism/antiferromagnetism  Phase Transition,¡± arXiv:1404.7737 [hep-th].

\bibitem{kondo}  Kondo, ``Resistance Minimum in Dilute Magnetic Alloys'', Prog. Theo. Phys. 32 (1964), no. 137¨C49.

\bibitem{holokondo1} P. Benincasa and A. Ramallo, ``Holographic Kondo Model in Various Dimensions'', JHEP 1206  (2012) 133, [arXiv:1204.6290].

\bibitem{holokondo2} J. Erdmenger, C. Hoyos, A. O'Bannon, and J. Wu,  ``Holographic Model of the Kondo  Effect'', JHEP 1312 (2013) 086, arXiv:1310.3271 [hep-th].

\bibitem{Garcia-Garcia:2013rha}
  A.~M.~Garc\'ia-Garc\'ia, H.~B.~Zeng and H.~Q.~Zhang,
  ``A thermal quench induces spatial inhomogeneities in a holographic superconductor,''
  JHEP {\bf 1407}, 096 (2014)
  [arXiv:1308.5398 [hep-th]].

\end{thebibliography}
\end{document}